\documentclass[doublecol]{epl2}
\usepackage{graphicx}
\usepackage{amssymb}
\usepackage{amsmath}
 
\usepackage{hyperref}
\hypersetup{
   bookmarks=true,         
    unicode=false,          
    pdfmenubar=true,        
    pdffitwindow=true,      
    pdftitle={My title},    
    pdfauthor={Author},     
   pdfsubject={Subject},   
    pdfcreator={Creator},   
    pdfproducer={Producer}, 
    pdfkeywords={keywords}, 
    pdfnewwindow=true,      
   colorlinks=true,       
   linkcolor=red,          
    citecolor=blue,        
    filecolor=magenta,      
    urlcolor=blue           
}

\DeclareMathAlphabet{\bi}{OML}{cmm}{b}{it}

\def\br{{\bf r}}

\def\b0{{\bf 0}}
\def\la{\langle}
\def\ra{\rangle}

\def\beq{\begin{equation}}
\def\eeq{\end{equation}}
\def\bea{\begin{eqnarray}}
\def\eea{\end{eqnarray}}
\def\bdm{\begin{displaymath}}
\def\edm{\end{displaymath}}

\def\e{\epsilon}

\title{Probing the optical conductivity of trapped charge-neutral quantum 
gases}
\author{Zhigang Wu\inst{1} \and Edward~Taylor \inst{2} \and Eugene~Zaremba\inst{1}}
\shortauthor{Z.~Wu \etal}

\institute{                    
  \inst{1} Department of Physics, Engineering Physics and Astronomy,
Queen's University, Kingston, Ontario, K7L 3N6, Canada\\
  \inst{2} Department of Physics and Astronomy, McMaster University, 
Hamilton, Ontario, L8S 4M1, Canada
}

\pacs{67.85.-d}{Ultracold gases, trapped gases}
\pacs{67.10.Jn}{Transport properties and hydrodynamics}
\pacs{03.75.Kk}{Dynamic properties of condensates; collective and hydrodynamic excitations, superfluid flow}

\date{\today}

\abstract{
We study a harmonically confined atomic gas which is subjected to an 
additional external potential such as an optical lattice. 
Using a linear response formulation, we determine the response of the gas 
to a small, time-dependent displacement of the harmonic 
trap and derive a simple exact relation showing that the 
centre-of-mass position of the atomic cloud is directly related to the 
global optical conductivity of the system. We demonstrate the 
usefulness of this approach by calculating the optical conductivity of bosonic atoms
in an optical lattice. In the Mott insulating phase, there is clear 
evidence of an optical Mott gap, providing a 
proof-of-principle demonstration that the global optical conductivity gives
high-quality information about the excitations of 
strongly-correlated quantum gases. }

\begin{document}

\maketitle
 
\section{Introduction}
A major goal in the field of ultracold 
atomic gases is to use these simple and tuneable systems to simulate 
strongly-correlated electronic systems in order to better understand 
the properties of the latter.  To this end, the field has begun to carry out 
the same kinds of measurements that have played a 
crucial role in unravelling the properties of electronic materials.  
These include high-precision thermodynamics~\cite{Ku12}, 
momentum-resolved radio-frequency spectroscopy~\cite{Stewart08} (the 
analogue of angle-resolved photo-emission spectroscopy), and transport 
measurements. The last includes conductance through a narrow channel 
separating two atomic gas reservoirs~\cite{ETH}, as well as transport 
quantities associated with \emph{charge-neutral} quantum fluids, 
namely viscosity~\cite{Cao11} and both the longitudinal~\cite{Sommer11} and 
transverse~\cite{Koschorreck13,Bardon13} spin diffusivity.

 On the other hand, experiments have yet to probe the analogue of the 
electrical conductivity $\sigma$ of quantum gases.  Conductivity 
measurements have played a major role in characterizing peculiar 
properties of charged quantum materials including the celebrated 
integer~\cite{vonKlitzing80} and fractional~\cite{Tsui82} quantum Hall 
effects, the mysterious ``strange metal''~\cite{Hussey08} and 
``pseudogap''~\cite{Timusk99} phases of the high-$T_c$ cuprate 
superconductors, as well as strongly-correlated (Mott) 
insulators~\cite{Thomas94}.  It would obviously be of great interest to 
be able to observe the analogue of conductivity in neutral quantum gases 
where novel quantum behaviour can be expected to occur with
increasing interaction strength. 
 
In this paper, we propose a way to probe the global optical conductivity 
in harmonically-confined gases and calculate this quantity for a Bose 
gas in a periodic optical potential.  We show that the response of the 
centre of mass of the gas to a periodic displacement of the harmonic 
trap directly yields information on the retarded correlation function 
for the \emph{total} current $\mathbf{J} = \int d\br 
\mathbf{j}(\br,t)$.  This opens the door to carrying out detailed 
measurements of the global optical conductivity tensor 
$\Sigma_{\alpha\beta}(\omega)$ for a range of systems, including bosonic 
Mott insulators~\cite{Greiner02}, the sought-after fermionic Mott 
insulator~\cite{Jordens08}, strongly-interacting fermions in an optical lattice, and 
fractional quantum Hall systems~\cite{Cooper13}.

Related ideas have been explored experimentally by McKay \textit{et 
al.}~\cite{McKay08} as well as theoretically by Tokuno and 
Giamarchi~\cite{Tokuno11} (see also, Ref.~\cite{Sensarma09}).   In the former, resistive dissipation was 
observed in the centre-of-mass dynamics of a Bose gas in an optical 
lattice potential after the trap was suddenly displaced, without resolving
the conductivity itself. Our work 
builds on this idea by considering a time-dependent periodic 
displacement of the trap potential. We show that detailed information
about the optical conductivity can be obtained by simply measuring the 
dynamics of the centre of mass of the cloud.  In contrast, Tokuno 
and Giamarchi~\cite{Tokuno11} emphasize the 
role played by a phase modulation of the optical lattice potential itself in the 
absence of harmonic confinement. In this case they show that the 
optical conductivity can be probed by measuring the rate of energy 
absorption in response to the phase modulation.

To give insight into what information can be inferred from a 
modulation of the harmonic trapping potential under realistic experimental 
conditions, we calculate the global optical conductivity 
for a Bose--Hubbard model in both the deep 
superfluid and the Mott insulator regimes, without resorting to a local 
density approximation. Despite the inhomogeneity of the trapped system, 
sharp signatures of quantum phases are clearly evident in the optical
conductivity. In the deep superfluid regime, the real part of the 
optical conductivity is dominated by a delta-function peak at the 
lowest excitation frequency, the analogue of the 
``Ferrell--Glover--Tinkham'' delta-function peak in 
superconductors~\cite{Ferrell58}. On the other hand, a distinct Mott gap 
appears in the spectrum of the optical conductivity in the Mott insulating 
regime. These examples show that the study of the centre-of-mass oscillations of 
trapped quantum gases can serve as high-quality simulators of the 
optical conductivity of strongly-correlated electronic materials, which 
are not confined in traps.

\section{Measuring the global conductivity}
In charged systems, the 
local conductivity $\sigma(\omega)$ is measured by applying a 
uniform electric field varying in time with frequency $\omega$.  
This generates a frequency-dependent current density $\textbf{j}$ in accordance with Ohm's 
law 
\beq j_{\alpha}(\omega) = \sum_\beta
\sigma_{\alpha\beta}(\omega)E_{\beta}(\omega)\label{Ohm}\eeq
(here and throughout this paper $\alpha$ and $\beta$ denote 
Cartesian components).   Microscopically, this leads to 
the Kubo expression 
\beq \sigma_{\alpha\beta}(\omega) = (ie^2/\omega 
V)[(N/m)\delta_{\alpha\beta} 
-\chi^J_{\alpha\beta}(\omega)],\label{sigmalocal}\eeq
for the local conductivity tensor, where $N$ is the total number of 
conducting electrons, $V$ is the volume of the system and 
$\chi^J_{\alpha\beta}(\omega)$ is the Fourier transform of the retarded 
current correlation function.  For a vector operator $\hat {\bf O}$, 
the correlation function of interest is defined by
\beq
\chi^O_{\alpha\beta}(t-t') \equiv \frac{i}{\hbar}\Theta(t-t')\langle 
[\hat{O}_{\alpha}(t),\hat{O}_{\beta}(t')]\rangle\label{chiO}.
\eeq

Ohm's law is not restricted to charged systems: it amounts to a 
statement about the current that arises in response to a force which, in the case of
a particle of charge $q$ subjected to an electric field $\mathbf{E}$, is $\mathbf{F}=q\mathbf{E}$.   
For a charge-neutral quantum gas confined in the harmonic trapping 
potential $V_{\mathrm{trap}} = \sum_{\alpha} 
m\omega^2_{\alpha}r^{2}_{\alpha}/2$, a small displacement of the trap from its equilibrium position generates a perturbing potential that is 
proportional to the displacement and hence, acts as a time-dependent uniform ``electric field". For a trap displacement $d_\beta (t)$ in the 
specific direction $\beta$, $\delta V_{\rm trap}(\br,t)= -m\omega_\beta^2 d_\beta(t)r_\beta$ (ignoring a dynamically irrelevant constant),
and the resulting spatially-independent force is ${\bf F}(t) \equiv  -\nabla\delta V_{\rm trap} \equiv m\omega_\beta^2 d_\beta (t)\hat {\bf x}_\beta$ .

The trap displacement results in the many-body perturbation $\delta \hat H(t) = -F_\beta(t) \sum_{i=1}^N \hat r_{i\beta} = -NF_\beta(t) \hat R_\beta$
where $\hat {\bf R} \equiv N^{-1} \sum_{i=1}^N \hat \br_i = N^{-1}\int d\br \,{\bf r} \hat{n}(\br)$ is the centre-of-mass co-ordinate of a cloud of $N$ atoms. For 
a \emph{small} displacement an application of linear response theory gives the cloud displacement
\beq
R_\alpha(\omega) = 
N\chi_{\alpha\beta}^{R}(\omega)F_\beta(\omega),
\label{Zre_fre}
\eeq
where $\chi_{\alpha\beta}^R(\omega)$ and $F_\beta(\omega)$ are the 
Fourier transforms of the retarded correlation function defined in (\ref{chiO}) for 
the centre-of-mass position $\hat {\bf R}$ of the trapped 
gas  and the force $F_\beta(t)$. The criterion for the validity of linear response theory depends on the physical context but can generally be taken as the condition
that the applied force $F_\beta(t)$ is small in comparison to the characteristic forces experienced by the particles in the system.

To make contact with the optical conductivity, 
we make use of the exact operator identity
\beq N\frac{ d\hat{R}_{\alpha}(t)}{dt} = 
\hat{J}_{\alpha}(t),\label{continuity}\eeq 
where $\hat{J}_{\alpha}(t) \equiv \int d\br \hat{j}_{\alpha}(\br,t)$ is 
the $\alpha$ component of the total current operator.  Combining this with (\ref{Zre_fre}) gives
\beq
J_\alpha (\omega) = -i N\omega R_\alpha(\omega) = -i N^2\omega 
\chi_{\alpha\beta}^{R}(\omega)F_\beta(\omega).\label{Ohmtrap}
\eeq
If we identify the global conductivity tensor as 
\beq
\Sigma_{\alpha\beta} (\omega) = -i N^2\omega 
\chi_{\alpha\beta}^{R}(\omega),
\label{Sig_def}
\eeq
(\ref{Ohmtrap}) is precisely Ohm's law.

The connection between (\ref{Sig_def}) and the usual Kubo formula in 
terms of the current correlation function follows from the conservation 
law (\ref{continuity}), which gives the exact relation 
$N^2\omega^2 \chi_{\alpha\beta}^{R}(\omega) =(iN/\hbar)\langle 
[\hat{R}_{\alpha}(0),\hat{J}_{\beta}(0)]\rangle +  
\chi_{\alpha\beta}^{J}(\omega)$.  Using this in (\ref{Sig_def}), the 
latter reduces to the standard-looking Kubo formula
\beq
\Sigma_{\alpha\beta}(\omega) = \frac{i}{\omega}\left 
\{-\frac{iN}{\hbar}\langle 
[\hat{R}_{\alpha}(0),\hat{J}_{\beta}(0)]\rangle -  
\chi_{\alpha\beta}^{J}(\omega)\right \}.
\label{Sig_kubo}
\eeq
The diamagnetic term given by the first term in curly brackets 
determines the sum rule for the frequency integral of the real 
conductivity; it is equal to $(N/m)\delta_{\alpha\beta}$ for 
any physical Hamiltonian, even when a trap is present.  However, for models 
with an implicit energy cutoff such as the Bose--Hubbard 
Hamiltonian we consider below, it assumes a different value 
[c.f. (\ref{fsumhb})].

If we take the trap displacement to be periodic, $d_{\beta}(t) 
=(d_{\beta}/2)\exp(-i\omega t ) + \mathrm{c.c.}$, the centre-of-mass position 
of the cloud is given by
\beq	 R_{\alpha}(t) = A_{\alpha}(\omega)\cos\left[\omega t 
-\phi_\alpha(\omega)\right],\label{Zw}\eeq
where the amplitude $A_{\alpha}(\omega)$ is the maximum displacement of the cloud from 
equilibrium and $\phi_{\alpha}(\omega)$ 
is the phase lag between the oscillation of the cloud and that of the potential.  
A combination of (\ref{Zre_fre}) and (\ref{Sig_def}) then gives the complex conductivity
\beq
\Sigma_{\alpha\beta}(\omega) = \frac{2N \omega}{iF_\beta}
A_{\alpha}(\omega)e^{i\phi_\alpha(\omega)}
\label{Sig_R_spe}
\eeq
which is thus completely determined by the dynamics of  the centre-of-mass position.

Equation (\ref{Sig_R_spe}) constitutes a major result of this 
paper.  It shows that the centre-of-mass  dynamics $R_\alpha(t)$ 
provides a measurement of the global optical conductivity tensor for a trapped gas, defined 
in (\ref{Sig_kubo}).   This tensor will be diagonal unless time-reversal symmetry is broken---as would happen in a rotating gas~\cite{Ho00,Ho01} or  by creating an artifiical gauge field~\cite{Lin09}---giving an off-diagonal Hall conductivity $\Sigma_{xy}\neq 0$.  As an illustration of the sort of information that is contained in the global conductivity, we calculate below $\Sigma_{xx}(\omega)$ for a 
one-dimensional trapped Bose-Hubbard model in both the deep superfluid and Mott insulating regimes, two paradigmatic states of quantum matter.

Before doing this, we note that the optical conductivity of a gas 
without any external potential other than the harmonic trap follows 
straightforwardly from the generalized Kohn theorem (see e.g., 
Ref.~\cite{Wu14}).  Because of a separation between the 
centre-of-mass and internal degrees of freedom, the optical conductivity is 
\beq
\Sigma_{\alpha\beta}(\omega)=\delta_{\alpha\beta}\frac{iN}{m}\frac{\omega}
{\omega^2-\omega_\alpha^2+i\eta},
\label{Kohn}
\eeq
where $\eta$ is a positive infinitesimal. This result is independent of interactions and shows
a resonant response at the trap frequency $\omega_\alpha$. In this case, the validity of 
linear response theory requires $d_\beta \ll  a_{{\rm ho},\beta}$, where $a_{{\rm ho},\beta} =\sqrt{\hbar/m\omega_\beta}$ 
is the relevant harmonic oscillator length.

\section{Global conductivity in a trapped Bose-Hubbard model}
The one-dimensional Bose--Hubbard model with harmonic confinement is~\cite{Jaksch98} 
\begin{align}
\hat H_{\rm BH}&=-t\sum_{j}\left (\hat a^\dag_{j}\hat a_{j+1} 
+\mathrm{h.c.}\right ) +\frac{1}{2}U\sum_j\hat n_j(\hat n_j-1) 
\nonumber \\
&\quad+\sum_j \e_j\hat n_j.
\label{HB_tr}
\end{align}
Here $t$ is the hopping matrix element, $U$ is the on-site repulsion and 
$\e_j=j^2 \e_0 $ describes the trapping potential with $\e_0 \equiv 
m\omega_0^2 a^2/2$, where $m$ is the mass of the particle, $\omega_0$ 
is the trap frequency and $a$ is the lattice spacing. If $\omega_{\rm opt}$ is the effective
oscillator frequency at a lattice site, the characteristic force on the particles arising from the optical lattice 
is $m\omega_{\rm opt}^2 a$. The criterion for the validity of linear response theory in this case is then
$(\omega_\beta/\omega_{\rm opt})^2(d_\beta/a) \ll 1$. Since $\omega_\beta \ll \omega_{\rm opt}$, linear response
theory can be valid even when $d_\beta \gg a$.

The 
centre-of-mass and the total current operators for this model  are 
$
\hat R = N^{-1}\sum_j aj\hat n_j
$, and 
$\hat J = (t a/i\hbar)\sum_j\left (\hat a^\dag_j\hat 
a_{j+1}-\mathrm{h.c.}\right )
$.   Using these to evaluate the diamagnetic term in (\ref{Sig_kubo}) gives 
the well-known sum rule 
\beq
\frac{1}{\pi}\int_{-\infty}^\infty d\omega  {\rm Re} 
\Sigma_{xx}(\omega) = -\frac{a^2}{\hbar^2}\la \hat T\ra,
\label{fsumhb}
\eeq
where $\hat T\equiv -t\sum_{j}(\hat{a}^{\dagger}_j\hat{a}_{j+1} + 
\mathrm{h.c.})$ is the kinetic energy operator in (\ref{HB_tr}).

To evaluate the real part of the zero-temperature ($T=0$) optical conductivity, we use 
(\ref{Sig_def})  in conjunction with the spectral representation 
\beq
{\rm Im} \chi^R(\omega)= \pi\sum_n|\la\Phi_0|\hat 
R|\Phi_n\ra|^2[\delta(\hbar\omega-E_{n0})-\delta(\hbar\omega+E_{n 0})],
\label{ImchiR}
\eeq
where $|\Phi_n\ra$ is an exact energy eigenstate of $\hat H_{\rm BH}$ 
and $E_{n 0}\equiv E_n-E_0$ is the excitation energy.  The real part follows from the Kramers--Kronig relations.  

We first consider the deep superfluid limit ($t\gg U$) 
where mean-field theory provides a good description of the ground state 
and low-energy excitations. The condensate wave function $\phi_j = \la 
\hat a_j\ra$, subject to the normalization $\sum_j|\phi_j|^2 = N$, is 
determined by the discrete Gross-Pitaevskii (GP) equation 
\begin{align}
-t(\phi_{j-1}+\phi_{j+1}) + (\epsilon_j + U|\phi_j|^2)\phi_j = \mu
\phi_j, 
\label{discrete_GP}
\end{align}
where 
$\mu$ is the chemical potential.   If the wave function amplitude $\phi_j$ varies slowly on the
scale of the lattice constant $a$, the substitution $\phi_j \to
\phi(x)$ can be used to convert (\ref{discrete_GP}) into the 
continuum GP equation
\beq -\frac{\hbar^2}{2m^* }\frac{\partial^2\phi}{\partial x^2} 
+ \left (\frac{1}{2}m^*\omega^{*2}x^2 + U\phi^2 \right ) \phi = \mu^*\phi,
\label{continuum_GP}
\eeq
where $\mu^* = \mu -2t$, $m^* = \hbar^2/2ta^2$ is the band mass and 
$\omega^*=\sqrt{m/m^*}\omega_0=2\sqrt{t\e_0}/\hbar$. 

Equation~(\ref{continuum_GP}) effectively describes a condensate
confined within a harmonic trap of frequency $\omega^*$ with the
particles having the renormalized mass $m^*$. The excitations of this system
are determined by the Bogoliubov-de Gennes (BdG) equations which yield
the amplitudes $u_n(x)$ and $v_n(x)$ with
the corresponding mode frequencies $\omega_n$.
Within the BdG approximation, we have
\beq
\la \Phi_0| \hat R |\Phi_n\ra  = \frac{1}{N}\int dx x 
\phi(x)\left [u_n(x)-v_n(x) \right ].
\label{Rmatr_con}
\eeq
In the appendix, we show that this quantity vanishes for all excited states 
except for the dipole 
mode for which~\cite{Fetter98}
\beq
\left ( u_{\rm dip} \atop v_{\rm dip} \right ) = 
\sqrt{\frac{m^*\omega^*}{{2N\hbar}}}\left(x\phi(x) \mp 
{\frac{\hbar}{m^*\omega^*}} \frac{d \phi(x)}{d x}\right ).
\label{uvdp}
\eeq
Substituting (\ref{uvdp}) into (\ref{Rmatr_con}) we thus obtain $ 
\la \Phi_0| \hat R |\Phi_{\rm dip}\ra  = 
\sqrt{{\hbar}/{2Nm^*\omega^*}}
$.
Using this result in (\ref{ImchiR}) and (\ref{Sig_def}), gives
\beq
\Sigma_{xx}(\omega) = i\frac{ N}{m^*}\frac{\omega}{\omega^2- 
\omega^{*2} + i\eta},
\label{Sig_super}
\eeq
which is analogous to (\ref{Kohn}). We note that this result for the optical conductivity deep in 
the superfluid regime obtained within the continuum GP description satisfies the $f$-sum rule 
(\ref{fsumhb})  with $\la\hat T\ra = -2tN=-N\hbar^2/m^*a^2$.

Moving away from the $U/t \to 0$ limit, the continuum GP approximation can no longer be invoked and the Kohn-like response given by (\ref{Sig_super}) does not  hold. Specifically,  we have confirmed by means of a numerical solution of the discrete BdG equations that, although the pole given by (\ref{Sig_super}) persists, its weight is reduced with increasing $U/t$. By analogy with the optical conductivity of superconductors~\cite{Ferrell58}, the weight of the pole provides a measure of the superfluid fraction  $N_s/N$; as a result of enhanced quantum fluctuations, $N_s/N < 1$ at $T=0$.  (In a superconductor, both the diamagnetic and paramagnetic current correlation functions contribute Dirac delta functions at $\omega = 0$ to the real part of the optical conductivity [c.f. (\ref{Sig_kubo})]; the net weight of the delta function contribution provides a definition of the superfluid fraction~\cite{Baymbook}.)

\begin{figure}
\onefigure[width=0.8\linewidth]{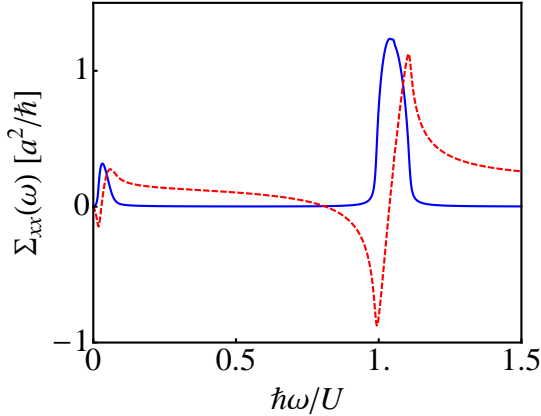}
\caption{Real (blue solid line) and imaginary (red dotted line) parts of the global
optical conductivity for a one-dimensional Bose-Hubbard model in a 
harmonic trap with $N=120$ atoms. The trap 
parameter $\e_0/U= 10^{-5}$. A Lorentzian broadening is introduced in (\ref{ImchiR}) with the replacement
$\delta(\hbar\omega - E_{n0})\to  \hbar \eta/[(\hbar\omega - E_{n0})^2+(\hbar\eta)^2]$, where $\hbar \eta/U = 0.005$.}
\label{Bulksig}
\end{figure}

\begin{figure}
\onefigure[width=0.8\linewidth]{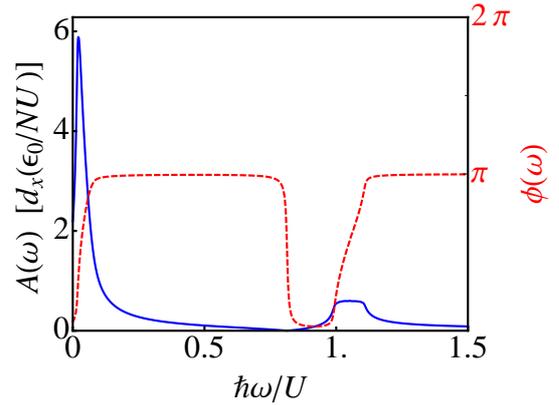}
\caption{The amplitude (blue solid line) and phase lag (red dashed line) of the centre of 
mass of the cloud  obtained from the global conductivity in 
Fig.~\ref{Bulksig}.}
\label{amp}
\end{figure}

We now consider the deep Mott-insulating regime ($ t\ll U$) with an occupancy $\bar n \simeq 1$ near 
the centre of the trap.   We perform numerical diagonalizations of 
(\ref{HB_tr}) using a truncated Hilbert space to obtain the ground and 
low-energy excited states of such an insulator.  The physically relevant states should 
include doublons, with energies of order $U$, as well as low-energy excitations involving atoms hopping to empty sites at the edges of the trap.  
Thus,  our truncated 
Hilbert space is spanned by the Mott-insulating ground state in the atomic limit, 
$
|\varphi_0\ra = \prod_{i=1}^{N} \hat a^\dag_{i} |0\ra$ (where the mid-point in the index range coincides with the trap centre), and the particle-hole excited states 
$|\varphi_{ji}\ra = \hat a^\dag_{j}\hat a_{i}|\varphi_0\ra
$ where the index $i$ is one of the occupied sites in $|\varphi_0\ra$ while 
$j$ is any possible occupied or empty site.  To allow for the latter possibility, we diagonalize (\ref{HB_tr}) using $N=120$ atoms and $M=150$ lattice sites.

The complex optical conductivity is shown in Fig.~\ref{Bulksig}  for $U/t=100$. 
Although this value is beyond what is typically explored in experiments (for the 3D experiment in Ref.~\cite{Greiner02}, the Mott-insulating transition was observed around $U/t = 36$ while in 1D, it will occur around $U/t=10$~\cite{Fisher89}), it ensures the accuracy of the approximations we have used in our numerical calculations.   We expect the behaviour to be qualitatively similar for smaller values $U/t$.   For frequencies $\omega\lesssim U/\hbar$, there is clear evidence of a Mott gap in the real part of the conductivity despite the presence of a harmonic trapping potential.  Such a feature is expected for untrapped Mott systems and arises from the absence of low-energy excitations in the Mott phase.  At higher frequencies $\omega\gtrsim U/\hbar$, doublons corresponding to double-occupancy states can be excited and the real part of the conductivity becomes significant, signifying dissipation.   The harmonic trapping potential manifests itself most directly through the presence of low-energy excitations corresponding to atoms being excited from the Mott insulator occupying the centre of the trap to adjacent vacant sites.  These give rise to the spectral feature at low frequencies in Fig.~\ref{Bulksig}.   

The cloud centre-of-mass dynamics corresponding to the conductivity in Fig.~\ref{Bulksig} is shown in Fig.~\ref{amp}.  Since the right-hand-side of (\ref{Sig_R_spe}) contains a factor $\omega$, the dynamics emphasizes the low-energy spectral weight in $\Sigma_{xx}(\omega)$.  For this reason, the role of the low-frequency edge excitations are greatly amplified in the centre-of-mass dynamics as compared to the high-frequency doublon contribution.  
The spectral features in the amplitude and phase associated with these modes can be understood in terms of damped, driven harmonic oscillators. The lower mode corresponding to the edge excitations has a frequency $\omega\sim t$ ($\sim 10^{-2}U/\hbar$ for our parameter values) which is associated with hopping from occupied to unoccupied sites. The ``Mott gap"  in which insulating behaviour is observed is then bounded by this frequency and the doublon frequency at $\omega\sim U/\hbar$. In this range of frequencies the cloud barely moves [$A(\omega)\simeq 0$] in response to the displacement. By the same token, the phase lag becomes $\pi$ since the cloud is being driven at frequencies above the natural resonant frequency of the lower mode [consistent with a purely imaginary conductivity, (\ref{Sig_R_spe})].  When the driving frequency approaches $U/\hbar$,  doublons are excited and the cloud once again begins to oscillate and $A(\omega)$ is substantial, corresponding to a nonzero real conductivity.  The large degeneracy of the doublons, on the order of the number of lattice sites, opens up a band of width $\gtrsim t$ where dissipation, as given by $\overline{dE/dt} = {\rm Re}\Sigma_{xx}(\omega)m^2\omega_0^4d^2_x/4$ (averaged over a period of oscillation), once again becomes significant. Above this band, the oscillation amplitude $A$ is small and the system is again effectively insulating. Since the driving frequency is above the doublon band, the oscillation of the cloud is out of phase with the driving field ($\phi\sim \pi$).

Although our results are for a 1D Mott insulator, the same primary features will carry over to 3D.  The major difference at high frequencies will be a wider doublon band~\cite{Sensarma09,Tokuno11}. At low frequencies, a pole in the conductivity will arise at $\omega^* = 2\sqrt{t\epsilon_0}/\hbar$ if the atoms at the edge of the cloud are superfluid since they can move around the Mott insulator in 3D.

\section{Summary}
In this paper we have shown using linear response theory that the centre-of-mass dynamics of an atomic cloud induced by an oscillating trapping potential is a sensitive probe of the optical conductivity of the gas.  As an illustration of the potential of conductivity measurements to obtain high-quality information about bulk states of matter, we calculated the optical conductivity of a Bose gas in the Mott insulating phase and the corresponding dynamics of the centre-of-mass of the cloud.  Despite the harmonic trap, the large separation $t\ll U$ of energy scales in the Mott insulator means that edge excitations particular to harmonically confined gases do not obscure clear signatures of a bulk Mott gap and doublon excitations.  Our scheme should be very useful in obtaining information about the properties of other states of matter, including integer~\cite{Ho00} and fractional~\cite{Cooper13} quantum Hall states, chiral $p$-wave superfluids~\cite{Levinsen11} via measurement of the off-diagonal response $\Sigma_{xy}$, as well as the diagonal spectral response of a strongly-interacting Fermi gas in a periodic potential, possibly connecting to spectral features of pseudogap physics~\cite{Timusk99}.  Finally, using a spin-selective perturbing potential, our results for the conductivity immediately carry over to the spin Hall conductivity, the central quantity that characterizes the spin Hall effect~\cite{Beeler13}.  

\acknowledgements{This work was supported by grants from the Natural Sciences and Engineering Research Council of Canada. } 

\section{Appendix--Dipole response of a harmonically confined gas in the BdG approximation}
Assuming a real ground state wavefunction $\phi(x)$, the BdG equations for excitations on top of the ground state described by the continuum GP equation (\ref{continuum_GP}) 
are
\beq \left[\begin{array}{cc} {\cal{L}} + 2g\phi^2 & -g\phi^2\\
-g\phi^2 & {\cal{L}} + 2g\phi^2\end{array}\right]\left[\begin{array}{c}u_n\\ v_n\end{array}\right] = \hbar\omega_n \left[\begin{array}{c}u_n\\ -v_n\end{array}\right],\label{BdG}\eeq
where
\beq {\cal{L}} \equiv -\frac{\hbar^2}{2m^* }\frac{\partial^2}{\partial x^2} +\frac{1}{2}m^*\omega^{*2}x^2 - \mu^*.\eeq
The spatially-varying Bogoliubov amplitudes $u_n(x)$ and $v_n(x)$ satisfy the orthonormality relations~\cite{Fetter72} 
\beq \int dx [u_n(x)u^{*}_m(x) - v_n(x)v^{*}_m(x)] = \delta_{nm}\label{normal}\eeq
and
\beq \int dx [u_n(x)v^*_m(x) - v_n(x)u^*_m(x)] = 0.\label{normal2}\eeq
Substitution of (\ref{uvdp}) in the main text into (\ref{BdG}) confirms that these are the Bogoliubov amplitudes of the dipole mode, with frequency $\omega_{\mathrm{dip}} = \omega^*$.  

As noted in Ref.~\cite{Fetter98}, the BdG equations can be formulated in terms of hydrodynamic variables corresponding to density $\delta \rho_n$ and phase $\theta_n$ fluctuations:
\beq 
{u_n \choose v_n} = \frac{i\phi \theta_n}{\hbar}\pm  \frac{\delta\rho_n}{2\phi}.\label{hydrovars}\eeq
Using this in (\ref{normal}) gives
\beq \int dx \delta \rho_n(x)\theta^*_m(x)\propto \delta_{nm}.\label{normal3}\eeq

The dipole mode corresponds to a ``rigid-body'' motion of the cloud, meaning that the velocity $v\propto \partial_x\theta$ is constant and hence, $\theta_{\mathrm{dip}}(x)\propto x$.  This is an exact result for systems obeying the generalized Kohn theorem~\cite{Wu14}, as is the case for the continuum GP model.  It can also be seen 
from (\ref{uvdp}) and (\ref{hydrovars}) that 
\beq \theta_{\mathrm{dip}}(x) = -i\sqrt{\frac{\hbar m^*\omega^*}{2N}}x.\eeq
Using this in (\ref{normal3}) gives
\beq  \int dx \delta \rho_n(x)x\propto \delta_{n,\mathrm{dip}}.\eeq
Using (\ref{hydrovars}) to rewrite $\delta\rho_n$ in terms of $u_n$ and $v_n$, this last result becomes
\beq \int dx x\phi(x) [u_n(x)-v_n(x)]\propto \delta_{n,\mathrm{dip}}.\eeq
This proves that the matrix element  given by (\ref{Rmatr_con}) in the main text vanishes for all excited states $n$ except for the dipole mode.

\end{document}